\documentclass[aps,prl,showpacs,superscriptaddress,
amssymb]{revtex4}

\usepackage{color}
\usepackage{graphicx}
\usepackage{dcolumn}
\usepackage{bm}
\usepackage{textcomp, pifont}

\begin{document}


\title{Anomalous Knight shift and low-energy spin dynamics in the nematic state of FeSe$_{\rm 1-x}$S$_{\rm x}$}

\author{V.\ Grinenko} \email{vadim.a.grinenko@gmail.com}
\affiliation{Institute for Solid State and Materials Physics, TU Dresden, 01069 Dresden, Germany}
\affiliation{IFW Dresden, Helmholtzstrasse 20, 01069 Dresden, Germany}
\author{S.\ Dengre}
\affiliation{Institute for Solid State and Materials Physics, TU Dresden, 01069 Dresden, Germany}
\author{R. Sarkar}
\affiliation{Institute for Solid State and Materials Physics, TU Dresden, 01069 Dresden, Germany}
\author{D. A. Chareev}
\affiliation{RAS, Institute of Experimental Mineralogy, Chernogolovka 123456, Russia}
\affiliation{Ural Federal University, Ekaterinburg 620002, Russia}
\affiliation{Kazan Federal University, Kazan 420008, Russia}
\author{A. N. Vasiliev} 
\affiliation{Lomonosov Moscow State University, Moscow 119991, Russia}
\affiliation{National University of Science and Technology "MISiS", Moscow 119049, Russia}
\affiliation{National Research South Ural State University, Chelyabinsk 454080, Russia}
\author{D. V.\ Efremov}
\affiliation{IFW Dresden, Helmholtzstrasse 20, 1069 Dresden, Germany}
\author{S.-L.\ Drechsler}
\affiliation{IFW Dresden, Helmholtzstrasse 20, 1069 Dresden, Germany}
\author{R.\ H\"uhne}
\affiliation{IFW Dresden, Helmholtzstrasse 20, 1069 Dresden, Germany}
\author{K.\ Nielsch}
\affiliation{IFW Dresden, Helmholtzstrasse 20, 1069 Dresden, Germany}
\author{H. Luetkens} 
\affiliation{Laboratory for Muon Spin Spectroscopy, PSI, CH-5232 Villigen PSI, Switzerland}
\author{H.-H. Klauss} 
\affiliation{Institute for Solid State and Materials Physics, TU Dresden, 01069 Dresden, Germany}

\date{\today}

\begin{abstract}
The interplay between the nematic order and magnetism in FeSe is not yet well understood. There is a controversy concerning the relationship between orbital and spin degrees of freedom in FeSe and their relevance for superconductivity. Here we investigate the effect of S substitution on the nematic transition temperature ($T_{\rm n}$) and the low-energy spin fluctuations (SF) in FeSe single crystals. We show that the low-energy SF emerge below the nematic transition. The difference between the onset temperature for the critical SF ($T_{\rm SF}$) and $T_{\rm n}$ is small for FeSe but significantly increases with S substitution. Below $T_{\rm SF}$ the Korringa relation is violated and the effective muon hyperfine coupling constant changes a sign. Our results exclude a direct coupling of the low-energy SF to the electronic nematic order indicating a presence of multiple spin degrees of freedom in FeSe$_{\rm 1-x}$S$_{\rm x}$.       
\end{abstract}


%
\maketitle 
One of the most puzzling  properties of FeSe is the nematic transition at $T_{\rm n}$ without a long-range magnetic order down to low temperatures.  To understand the nature of this state a number of theoretical and experimental approaches have been applied \cite{Onari2012, Fernandes2014, Glasbrenner2014, Yu2015, Wang2015, Yamakawa2016, Xing2017, Wang2016c, Lai2017, Kreisel2018, Hirschfeld2015, Chubukov2016}. It is generally accepted that the orbital and spin degrees of freedom are coupled in this system. Therefore, independent of the nature of the nematic transition both orbital and spin fluctuations (SF) should be affected at $T_{\rm n}$.  Indeed, nuclear magnetic resonance (NMR) and inelastic neutron scattering (INS) experiments on FeSe show that the nematic transition affects the SF spectra. The relaxation rate $1/T_1T$ increases below $T_{\rm n}$ \cite{Baek2015, Bohmer2015, Kasahara2016, Shi2018}. In turn, the INS measurements revealed that the SF at the antiferromagnetic (AF) vector $Q_{\rm N}$= [$\pi$, $\pi$] weaken with the reduction of the temperature and a stripe SF at $Q_{\rm S}$= [$\pi$ ,$0$]  enhances below $T_{\rm n}$ \cite{Wang2016a, Wang2016b}. Taking into account that the essential change of the SF is observed close to $T_{\rm n}$ the authors concluded that the SF spectra is directly affected by the nematic transition. Here, we show that any direct relationship between the nematic transition at $T_{\rm n}$ and the onset temperature $T_{\rm SF}$ of the low-energy SF is questionable in FeSe$_{\rm 1-x}$S$_{\rm x}$. These two temperatures are similar, only for stoichiometric FeSe but split significantly with S substitution.

High-quality FeSe$_{\rm 1-x}$S$_{\rm x}$ single crystals were synthesized in  eutectic molten metal chlorides under a permanent temperature gradient as described in Ref.\ \cite{Chareev2018}. The composition of the crystals was defined by microprobe analysis. The magnetization measurements were performed using a commercial superconducting (SC) quantum interference device (SQUID) magnetometer from Quantum Design. The specific heat and electrical transport were measured in a Quantum Design physical property measurement system (PPMS). $^{77}$Se-NMR experiments were performed using a Tecmag spectrometer in a magnetic field $B$ = 6.5T applied along the $c$-axis and in the $ab$-plane. The high-field muon spin rotation/relaxation ($\mu$SR) measurements were performed at the HAL-9500 spectrometer (PSI, Villigen) on an assembly of single crystals with a total sample mass of about 30 mg in a magnetic field $B$ = 9.5 T applied along the $c$-axis and in the $ab$-plane. To increase the amount of muons stopping in the sample, we used an Ag degrader with the thickness of $d_{\rm Ag} = $25 $\mu$m as described in Ref.\ \onlinecite{Grinenko2018, Grinenko2017}. The $\mu$SR data were analyzed using the musrfit software package.\cite{Suter2012}.

\begin{figure*}[ht]
\includegraphics[width=1\textwidth]{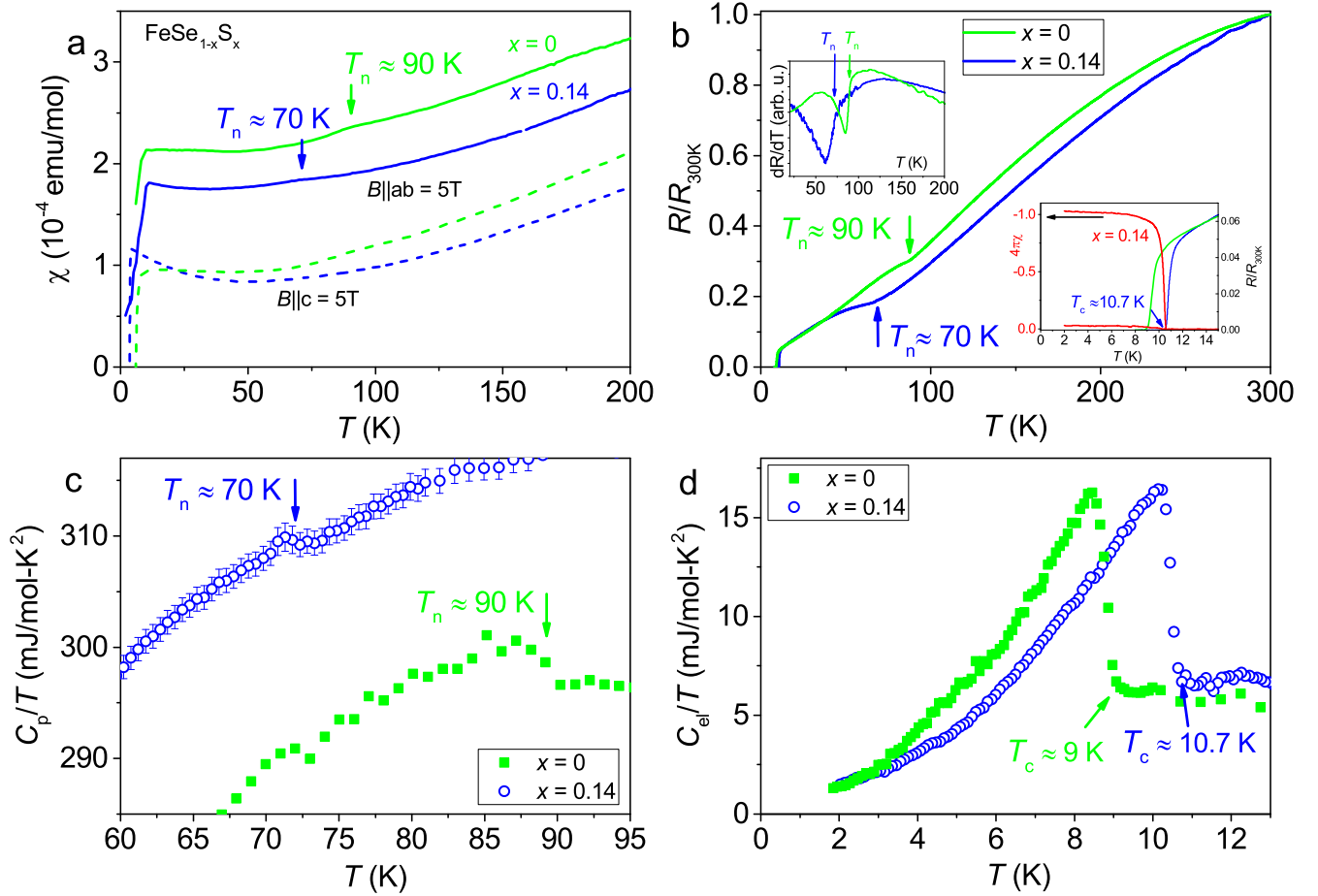}
\caption{(Color online) a) Temperature dependence of the molar magnetic susceptibility of the FeSe$_{\rm 1-x}$S$_{\rm x}$ samples used in the $\mu$SR experiments. b) Temperature dependence of the normalized electrical resistivity of the FeSe$_{\rm 1-x}$S$_{\rm x}$ single crystals measured in zero field. Upper inset shows the temperature dependence of the first temperature derivative of the electrical resistivity for the same crystals. Bottom inset shows the temperature dependence of the normalized electrical resistivity (right axes)  measured in zero field and the volume magnetic susceptibility (left axes) measured in $B\parallel ab$ = 0.5 mT close to the SC transition. c) Temperature dependence of the specific heat for the same crystals as shown in Fig. \ref{Fig:1}b close to the nematic transition temperature measured in zero field. d) Temperature dependence of the electronic specific heat for the same crystals as shown in Fig. \ref{Fig:1}c close to the SC transition.} 
\label{Fig:1}
\end{figure*}

All investigated crystals show a sharp SC transition (see Fig.\ \ref{Fig:1}b, \ref{Fig:1}d and Fig.\ S1 in the Supplementary materials (SM)). The nematic transition temperature $T_{\rm n}$ was obtained from the temperature dependencies of the magnetic susceptibility, electrical resistivity and specific heat as shown in Figs.\ref{Fig:1}a, \ref{Fig:1}b, and \ref{Fig:1}c, correspondingly. The $T_{\rm n}$ values measured by the different methods are in a good agreement between each other. $T_{\rm n}$ decreases with the substitution from $T_{\rm n} \approx 90$K for FeSe to $T_{\rm n} \approx 70$K for $x\approx 0.14$, where $T_{\rm c}$ increases from $T_{\rm c} \approx 9$K to $T_{\rm c} \approx 11$K, correspondingly. The $T_{\rm n}$ values of our samples for a given substitution level are slightly higher than those published in literature \cite{Wiecki2018, Matsuura2017}, which is, presumably, related to the application of the different methods for determining the substitution level. The discrepancy is eliminated if we use the crystallographic lattice parameters instead of the S substitution level $x$ for the comparison between our data and samples from the literature (see Fig.\ S2 in the SM).

\begin{figure*}[ht]
\includegraphics[width=1\textwidth]{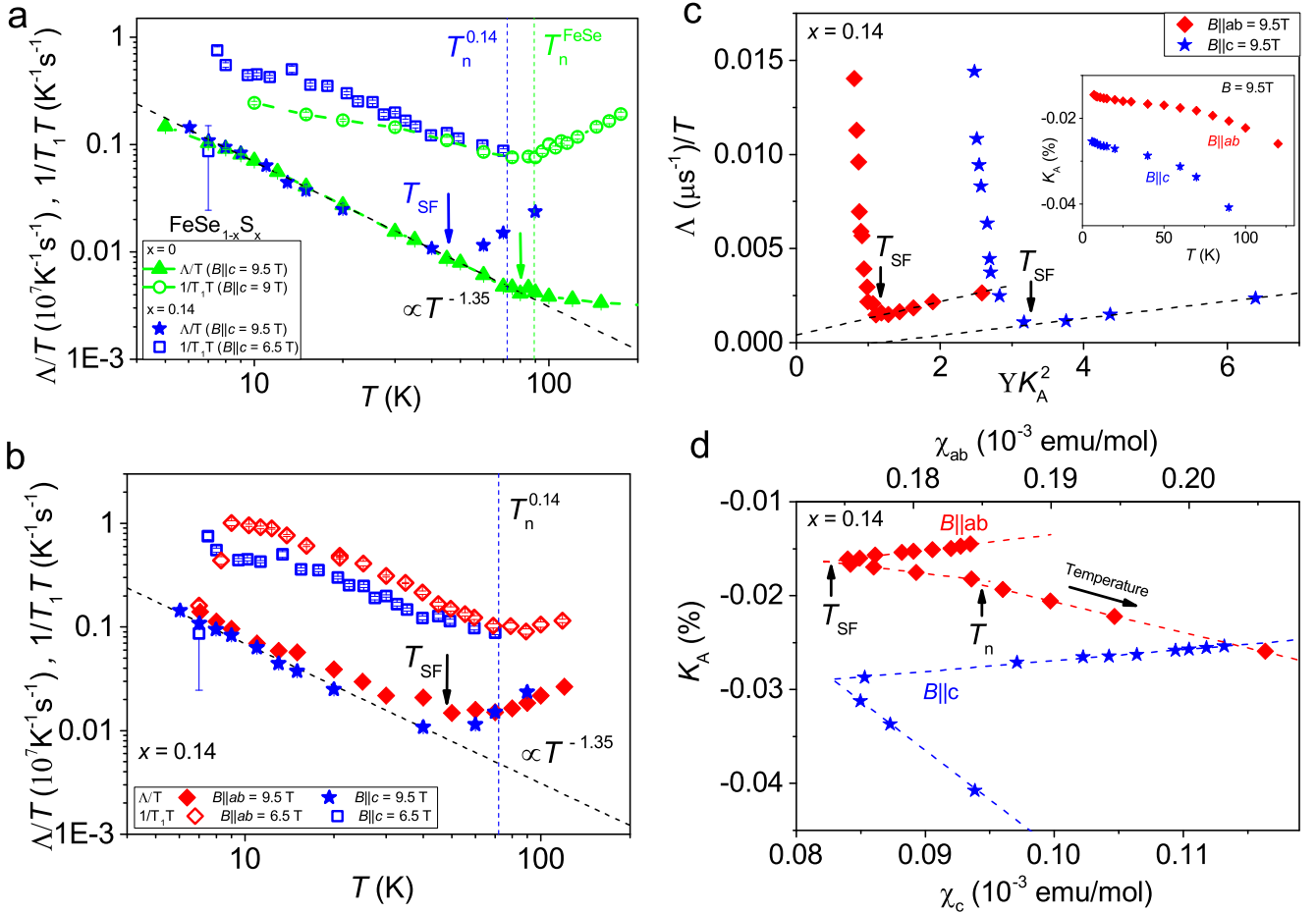}
\caption{(Color online) a) Log-log plot of the temperature dependence of the muon depolarization rate $\Lambda_{\rm TF}/T$ and the NMR spin-lattice relaxation rate $1/T_1T$ for the FeSe$_{\rm 1-x}$S$_{\rm x}$ single crystals in magnetic fields applied along the $c$-axes. The data for FeSe are taken from Refs. \cite{Grinenko2018, Bohmer2015}. b) Log-log plot for $\Lambda_{\rm TF}/T$ and NMR $1/T_1T$ of the single crystals with $x$ = 0.14 measured in magnetic field applied along two different crystallographic directions (muon spin polarization is perpendicular to the direction of the magnetic field). c) Korringa relation between $\Lambda_{\rm TF}/T$ and $\mu$SR Knight shift $K_{\rm A}$. Inset shows the temperature dependence of $K_{\rm A}$ (corrected for demagnetization effects) for the sample with $x$ = 0.14. d) $K_{\rm A}$  vs. bulk molar susceptibility $\chi$ for the sample with $x$ = 0.14 (the Clogston-Jaccarino plot). Lines are linear fits of the data.} 
\label{Fig:2}
\end{figure*}

The low-energy SF were investigated by NMR and $\mu$SR techniques. The temperature dependence of the spin-lattice relaxation rate $1/T_1$ for $^{\rm 77}$Se nucleus and the muon depolarization rate $\Lambda_{\rm TF}$ for $x = 0$ and 0.14 are compared in Fig.\ \ref{Fig:2}a. The $\Lambda_{\rm TF}$ values were obtained by fitting the transversal field spectra in a time domain as described in Ref. \onlinecite{Grinenko2018}. Examples of a Fast Fourier transform of the time spectra are given in the SM Fig.\ S3. There are several possible sources of the muon spins depolarization such as nuclear moments, electronic spins and inhomogeneous fields due to possible magnetic impurities. The nuclear contribution in the case of FeSe is rather weak \cite{Grinenko2018}. Also, our samples didn't show noticeable paramagnetic impurity contribution in the static susceptibility and the Knight shift ($K_{\rm A}$) (see Figs. \ref{Fig:1}a and \ref{Fig:2}c). Therefore, we expect that the muon relaxation rate is dominated by the interaction with electronic moments and hence $\Lambda_{\rm TF}\propto 1/T_1$.  At high temperatures both $1/T_1T$ and $\Lambda_{\rm TF}/T$ decrease while reducing the temperature (Figs.\ \ref{Fig:2}a and \ref{Fig:2}b). The reduction is attributed to the suppression of the N\'{e}el SF observed by INS \cite{Wang2016a}. In this temperature range we found that $\Lambda_{\rm TF}/T \propto K_{\rm A}^2$ as shown in Fig.\ \ref{Fig:2}c with the proportionally constant $Y = \frac{4 \pi k_{\rm B}}{\hbar}(\frac{\gamma_{\rm \mu}}{\gamma_{\rm el}})^2$, where $k_{\rm B}$ and $\hbar$ are Boltzmann and Plank constants, $\gamma_{\rm \mu}$ and $\gamma_{\rm el}$ are gyromagnetic ratios of muon and electron, respectively. Below $T_{\rm SF}$ both relaxation rates change the behavior indicating a strong enchantment of the low-energy SF. In FeSe this behavior was attributed to the enhancement of the critical stripe AF SF at low temperatures \cite{Grinenko2018, Wang2016a, Wang2016b}. At the same temperature the Korringa relation violates indicating a strong enhancement of the correlations (Fig.\ \ref{Fig:2}c).  
 
In spite of the similar trend in the temperature dependencies of $\Lambda_{\rm TF}$ and $^{\rm 77}$Se-NMR $1/T_1$ the expected proportionality between them ($\Lambda_{\rm TF}\propto 1/T_1$) is lacking. In particular, the critical behavior at low temperatures $\Lambda_{\rm TF} \propto T^{\rm -n}$ can be hardly found in the temperature dependence of $1/T_1$  (Figs.\ \ref{Fig:2}a and  \ref{Fig:2}b). 
To understated this discrepancy we considered the Clogston-Jaccarino plot shown in Fig.\ \ref{Fig:2}d 
for the $\mu$SR Knight shift 
$K_{\rm A}$ (the Clogston-Jaccarino plot for FeSe including the $^{\rm 77}$Se-NMR Knight shift 
($K^{\rm Se}$) can be found in Ref.\ \cite{Grinenko2018}). At low temperatures
$K_{\rm A}$ is a linear with $\chi$ indicating that coupling constant $A^{\mu}$ is temperature independent.  The linear relation is violated at $T_{\rm SF}$ and $A^{\mu}$ changes a sign above $T_{\rm SF}$, whereas the absolute $A^{\mu}$ value is nearly unchanged up to $T_{\rm n}$. Finally, about 10$\%$ change of the $A^{\mu}$ value is observed across the structural transition at $T_{\rm n}$ (Fig.\ \ref{Fig:2}d). The linear relationship between the NMR $K^{\rm Se}$ and the susceptibility is also violated but the Se hyperfine coupling constant $A_{\rm hf}^{\rm Se}$ doesn't change a sign \cite{Grinenko2018}. Instead, $K^{\rm Se}$ is nearly temperature independent at low temperatures, which can be interpreted as a strong reduction of $A_{\rm hf}^{\rm Se}$. The reduction of $A_{\rm hf}^{\rm Se}$ explains phenomenologically the lacking proportionality $\Lambda_{\rm TF}\propto 1/T_1$ at low temperatures.

The anomalous Clogston-Jaccarino plot (Fig.\ \ref{Fig:2}d) mimics the behavior found in heavy fermion compounds \cite{Shirer2012}. One of the possible microscopic interpretation is a coexistence of multiple spin degrees of freedom. The two spin components appear in Kondo lattice materials having both localized $f$ electrons and itinerant conduction electrons \cite{Shirer2012}. By the analogy with heavy fermion systems we propose that the anomalous behavior of the FeSe system can be explained by the coexistence of localized and itinerant electrons having different temperature dependencies and anisotropies of the related susceptibilities $\chi_{\rm loc}$ and $\chi_{\rm cond}$. Microscopically the localized and the itinerant electrons may originate from the Fe $3d$ orbitals due to a strong orbital selective correlation effect as proposed in Ref. \cite{Yu2018}. In this case, the muon Knight shift consists of dipolar ($K_{\rm dip}$), hyperfine ($K_{\rm hyp}$) and Fermi contact ($K_{\rm Fermi}$) terms, originating from dipole, and hyperfine fields  and interaction with conducting electrons, correspondingly. It is expected that in the case of local moments the total Knight shift $K_{\rm A}$ is dominated by dipolar and hyperfine terms, whereas in a pure itinerant system $K_{\rm Fermi}$ dominates. For an intermediate case all terms can have comparable values.  For the axially symmetrical muon site A ($T > T_{\rm n}$) by the first approximation one expect that the total Knight shift can be divided in two contributions arising from local and itinerant (conducting) electrons: $K_{\rm A}^{\rm ab} \approx (A_{\rm c}^{\rm ab} - \frac {1}{2}A_{\rm dip}^{zz})\chi_{\rm loc}^{\rm ab}+A_{\rm c}^{\rm ab}\chi_{\rm cond}^{\rm ab}$ for the magnetic field applied along the $ab$-plane and $K_{\rm A}^{\rm c} \approx (A_{\rm c}^{\rm c} + K_{\rm dip}^{zz})\chi_{\rm loc}^{\rm c}+A_{\rm c}^{\rm c}\chi_{\rm cond}^{\rm c}$ for the magnetic field applied along the $c$-axis.  Thus, a dramatic change of the $K_{\rm A}$ vs. $\chi$ dependence at $T_{\rm SF}$ at the same time a very weak effect across the structural transition at $T_{\rm n}$ can be explained within this model if the susceptibilities $\chi_{\rm loc}$ and $\chi_{\rm cond}$ have different temperature dependencies.

\begin{figure}[t]
\includegraphics[width=0.5\textwidth]{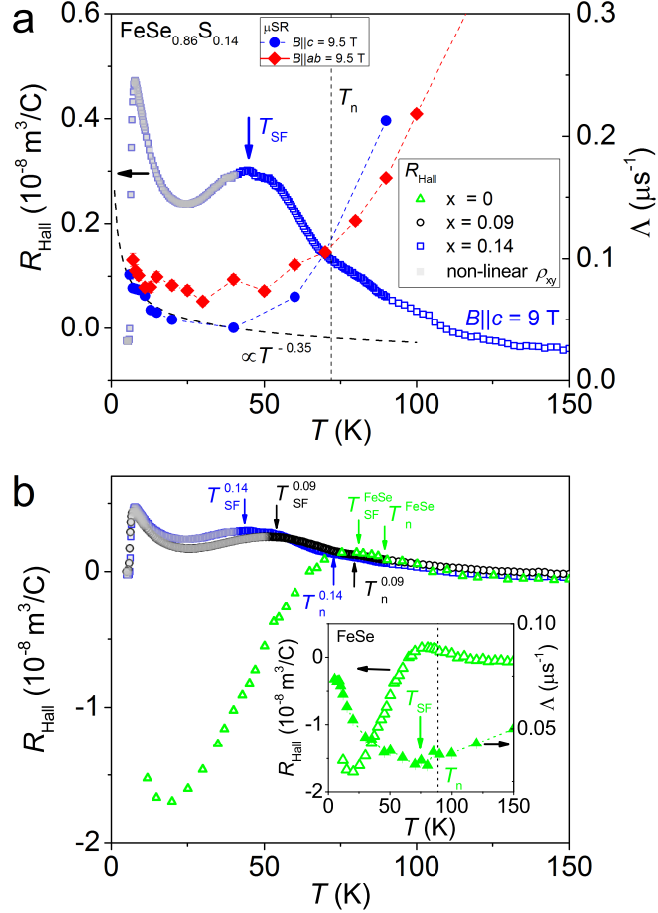}
\caption{(Color online) a) Temperature dependence of the Hall coefficient (left axis) and the muon depolarization rate $\Lambda_{\rm TF}$ (right axis) for FeSe$_{\rm 0.86}$S$_{\rm 0.14}$. b) Temperature dependence of the Hall coefficient for FeSe$_{\rm 1-x}$S$_{\rm x}$ single crystals measured in applied magnetic field B $\parallel c$ = 9 T. Inset shows temperature dependence of the Hall coefficient (left axis) and the muon depolarization rate $\Lambda_{\rm TF}$ (right axis) of FeSe. The data for FeSe are taken from Refs.\ \cite{Rossler2015} and \cite{Grinenko2018}.}
\label{Fig:3}
\end{figure} 

The anomaly at $T_{\rm SF}$ is  seen, also in the transport properties. We observed that the onset of the low-energy SF has a strong effect on the temperature dependence of the Hall coefficient (Fig.\ \ref{Fig:3}). The voltage $V_{\rm xy}$ is linear with the applied magnetic field $B {\rm \parallel} c$ along the crystal $c$-axis above $T_{\rm SF}$ in accord with previous studies \cite{Sun2016a,Ovchenkov2017} and acquires a non-linear contribution below $T_{\rm SF}$. Therefore, above $T_{\rm SF}$ the Hall coefficient can be defined as $R_{\rm Hall} = \rho_{\rm xy}/B_{\rm \parallel c} = (\rho(B_{\rm \parallel c})-\rho(-B_{\rm \parallel c}))/2B_{\rm \parallel c}$.  We found that the position of the maximum in $R_{\rm Hall}$ is very close to $T_{\rm SF}$ seen by $\mu$SR and therefore, allows to track a substitution dependence of $T_{\rm SF}$. Temperature dependencies of $R_{\rm Hall}$ for $x =$ 0 , 0.09, and 0.14 are shown in Fig.\ \ref{Fig:3}b. It is seen that the splitting between $T_{\rm n}$ and $T_{\rm SF}$ gradually increases with S substitution. For FeSe$_{\rm 0.86}$S$_{\rm 0.14}$ we observed that $T_{\rm SF}$ is about 30 K below $T_{\rm n}$.
 
Previously, the onset temperature of the stripe SF $T_{\rm SF}$ was related to the nematic transition since both $T_{\rm SF}$ and $T_{\rm n}$ are very close for a stoichiometric FeSe. It was concluded that the nematic transition is primary responsible for the changes in the SF spectra \cite{Wang2016a}. This direct coupling between orbital and spin degrees of freedom is an important fundamental observation and favors the itinerant SF scenarios for FeSe \cite{Kreisel2018, Hirschfeld2015, Chubukov2016}. Within this scenario one would expect that a reduction of $T_{\rm n}$ (in our case with S substitution) results in a similar or a somewhat slower suppression of $T_{\rm SF}$. However, in contrast to this expectation we found that $T_{\rm SF}$ is suppressed with S substitution faster than $T_{\rm n}$, as shown in Fig.\ \ref{Fig:2} and Fig.\ \ref{Fig:3}.  This in addition to the anomalous Clogston-Jaccarino plot indicates a presence of multiple spin and orbital degrees of freedom in the FeSe$_{\rm 1-x}$S$_{\rm x}$ system. We propose that the  orbital selective correlations are responsible for the complex behavior. It is predicted that the correlation effect depends on the temperature and can be enhanced by the nematic order \cite{Yu2018}. The enhancement depends on the interplay between different types of the neamtic order.  Therefore, a strength of the correlation effects for different Fe $3d$ orbitals may have a complex temperature dependence resulting in enhancement of the low energy SF at $T_{\rm SF}$ quite different from $T_{\rm n}$.    

In conclusion, we observed that the nematic transition at $T_{\rm n}$ does not affect directly the low-energy SF in FeSe$_{\rm 1-x}$S$_{\rm x}$. The splitting between  $T_{\rm n}$ and the onset temperature of the critical SF, $T_{\rm SF}$, increases with S substitution from $T_{\rm n}-T_{\rm SF}\sim$ 10 K for FeSe to $T_{\rm n}-T_{\rm SF}\sim$ 30 K for $x$ = 0.14. The region with critical SF shrinks with $x$ indicating that S substitution moves the system away from the quantum critical point in accord with transport data \cite{Ovchenkov2018}. In general, our results indicate that FeSe$_{\rm 1-x}$S$_{\rm x}$ shows a multicomponent behavior with rather decoupled spin degrees of freedom responsible for the low-energy SF and the nematicity. We attribute this behavior to the orbital selectivity of the electron correlations in the FeSe system. We believe that our work will stimulate further experimental and theoretical studies to understand this complex behavior.        

This work was supported by the DFG through grant GR 4667 and within the research training group GRK 1621. R.S. and H.H.K. are thankful to DFG for the financial assistance through the SFB 1143 for the project C02. This work was partially performed at Swiss Muon Source (S$\mu$S), PSI, Villigen. A.N.V. acknowledges the support from the Ministry of Education and Science of the Russian Federation in the framework of Increase Competitiveness Program of NUST 'MISiS' Grant No. K2-2017-084; by Act 211 of the Government of Russian Federation, Contracts No. 02.A03.21.0004, 02.A03.21.0006 and No. 02.A03.21.0011; the support from Russian Foundation for Basic Research, Grant No. 17-29-10007. D.A.C. thanks supports by the program 211 of the Russian Federation Government (RFG), agreement 02.A03.21.0006 and by the RFG Program of Competitive Growth of KFU. We acknowledge  fruitful discussions with A.\ Amato, A. Charnukha, T. Goko, K. Nenkov, R. Scheuermann, and Q.\ Si.

\newpage
\section{Supplementary material}

\renewcommand{\theequation}{S\arabic{equation}}
\renewcommand{\thefigure}{S\arabic{figure}}
\renewcommand{\thetable}{S\arabic{table}}
\setcounter{equation}{0}
\setcounter{figure}{0}
\setcounter{table}{0}

{In this supplementary material we provide additional magnetization data, and examples of the Fast Fourier transform (FFT)  of the $\mu$SR transversal field time spectra for the FeSe$_{\rm 1-x}$S$_{\rm x}$ single crystals. We also compare the substitution dependencies of the lattice parameters for our crystals with the literature data. }


\begin{figure}[b]
\includegraphics[width=30pc,clip]{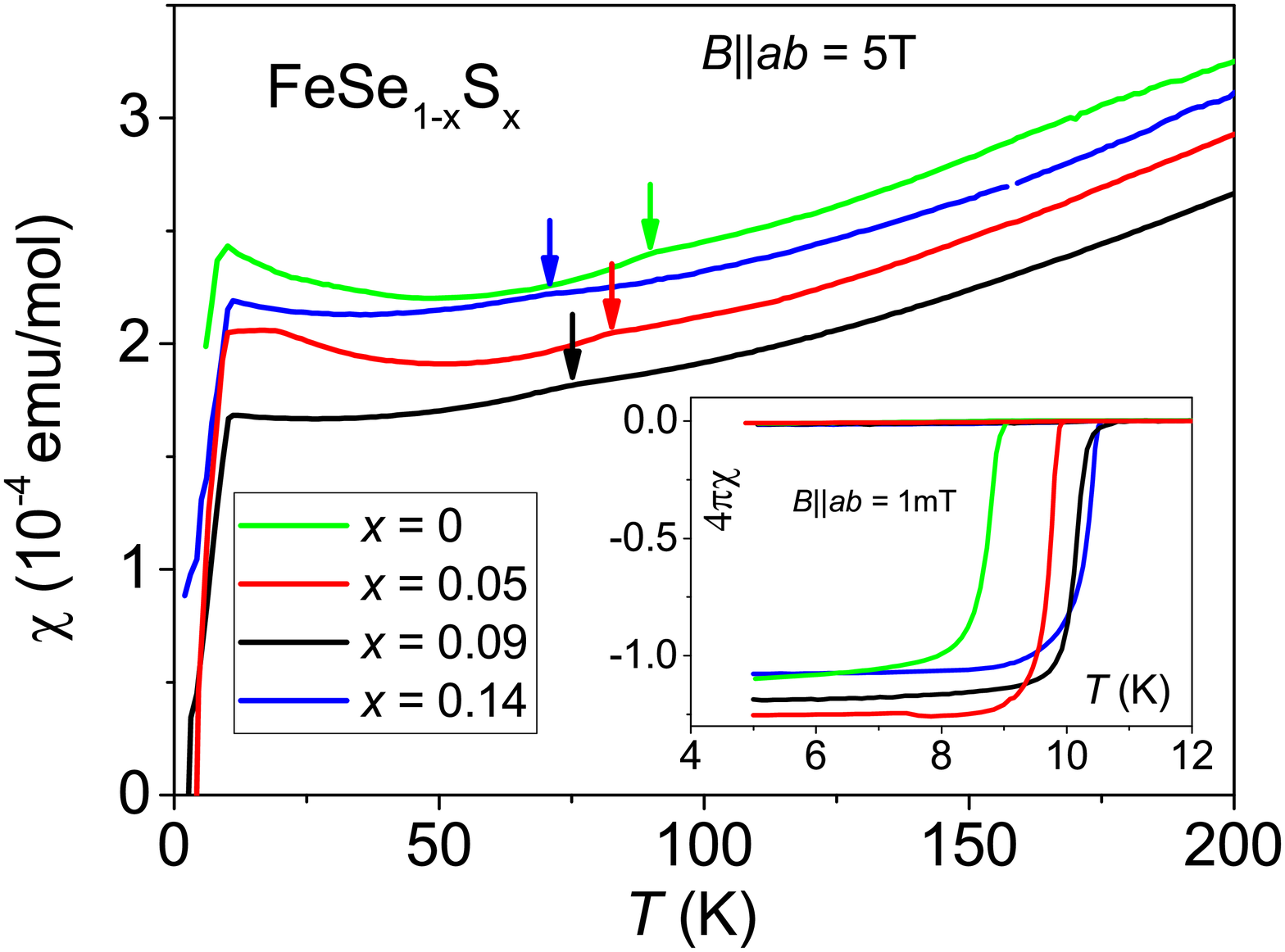}
\caption{Temperature dependence of the molar magnetic susceptibility for the FeSe$_{\rm 1-x}$S$_{\rm x}$ samples with different doing levels. The inset shows the temperature dependence of the volume susceptibility for the same crystals in the superconducting state.} 
\label{FigS:1}
\end{figure}


\begin{figure}[t]
\includegraphics[width=30pc,clip]{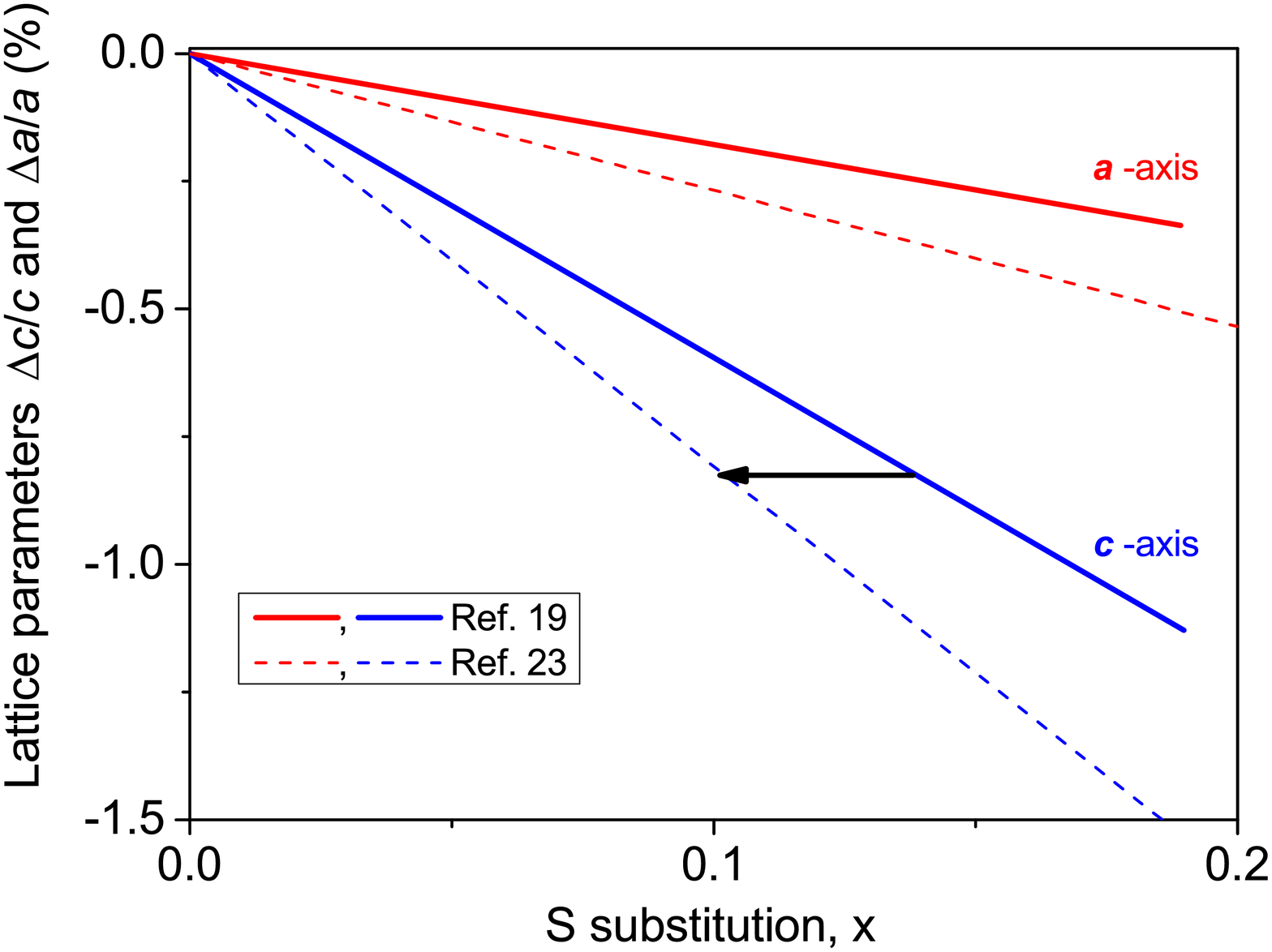}
\caption{The substitution dependence of the lattice parameters for FeSe$_{\rm 1-x}$S$_{\rm x}$ single crystals. Solid lines - the crystals prepared by the same method \cite{Chareev2018} as in our work. Dashed lines - the crystals from Ref. \cite{Wiecki2018}. The arrow shows the value for our crystal with a substitution level $x = 0.14$, which correspond roughly to $x = 0.1$ in the literature data.} 
\label{FigS:2}
\end{figure}

\begin{figure}[b]
\includegraphics[width=30pc,clip]{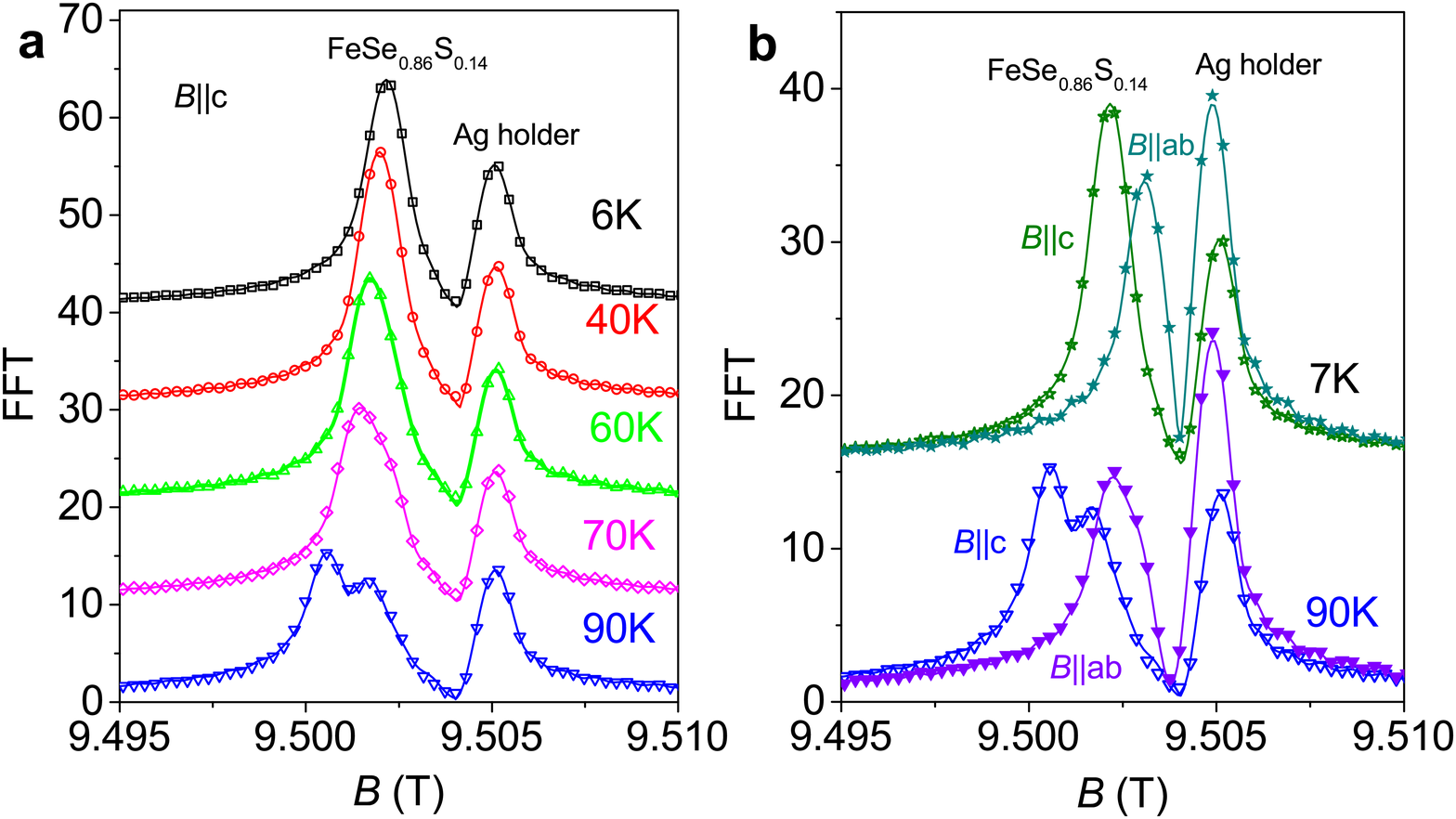}
\caption{a) FFT of the high transversal field $\mu$SR time spectra of FeSe$_{\rm 1-x}$S$_{\rm x}$ samples measured at different temperatures. b) Examples of the FFT obtained with the magnetic field applied along two different crystallographic directions (the muon spin polarization is perpendicular to the field direction).} 
\label{FigS:3} 
\end{figure}

\end{document}